\documentclass[3p,times,procedia]{elsarticle}
\usepackage{nupha_ecrc}

\volume{00}

\firstpage{1}

\journalname{Nuclear Physics A}

\runauth{V. Vislavicius}
\jid{nupha}
\jnltitlelogo{Nuclear Physics A}
\usepackage{amssymb}
\usepackage{amsmath}
\usepackage{xcolor}
\usepackage{xspace}
\usepackage[figuresright]{rotating}

\def\s#1{\ensuremath{\!\sqrt{s}~=~#1\textrm{~TeV}}}
\def\so{\ensuremath{\!\sqrt{s}}}
\def\pT{\ensuremath{p_{\textrm{\tiny T}}}}
\def\mpt{\ensuremath{\langle\pT\rangle}}

\def\fig#1{Fig.~\ref{fig:#1}}
\def\dndeta{\ensuremath{\langle\textrm{d}N_{\textrm{\tiny ch}}/\textrm{d}\eta\rangle}}
\newcommand{\overbar}[1]{\mkern 2mu\overline{\mkern-2mu#1\mkern-2mu}\mkern 2mu}
\begin{document}

\begin{frontmatter}

\dochead{XXVIth International Conference on Ultrarelativistic Nucleus-Nucleus Collisions\\ (Quark Matter 2017)}
\title{Multiplicity dependence of identified particle production in proton-proton collisions with ALICE}

\author{Vytautas Vislavicius on behalf of the ALICE collaboration}
\address{Lund University, Sweden}

\begin{abstract}
The study of identified particle production as a function of transverse momentum (\pT) and event multiplicity in proton-proton (pp) collisions at different center-of-mass energies (\so) is a key tool for understanding similarities and differences between small and large collisions systems. We report on the production of $\pi^{\pm}$, $K^{\pm}$, $K^{0}_{S}$, $p (\overbar{p})$, $\Lambda (\overbar{\Lambda})$, $\Xi^{\pm}$ and $\Omega^{\pm}$ measured in pp collisions in a wide range of center-of-mass energies with ALICE~\cite{Aamodt:2008zz}.
The multiplicity dependence of identified particle yields is presented for \so\ = 7 and 13~TeV and discussed in the context of the results obtained in proton-lead (p-Pb) and lead-lead (Pb-Pb) collisions, unveiling remarkable and intriguing similarities.
The production rates of strange hadrons are observed to increase more than those of non-strange particles, showing an enhancement pattern with multiplicity which does not depend on the collision energy.
Even if the multiplicity dependence of spectral shapes can be qualitatively described by commonly-used Monte Carlo (MC) event generators, the evolution of integrated yield ratios is poorly described by these models. 
\end{abstract}

\begin{keyword}
Multiplicity dependence, collectivity, small systems
\end{keyword}

\end{frontmatter}


\section{Introduction}
\label{intro}

Measurements of hadron yields as a function of multiplicity in p-Pb collisions at \s{5.02} revealed trends reminiscent to those observed in Pb-Pb collisions~\cite{Abelev:2013haa} and usually associated with the creation of a strongly interacting medium, the Quark-Gluon Plasma (QGP). Even more remarkably, a similar behavior was observed for particle production in high multiplicity pp collisions~\cite{Adam:2016emw}. Features like baryon-to-meson ratio enhancement at intermediate transverse momentum (\pT) in Pb-Pb collisions are understood as a consequence of quark coalescence~\cite{Fries:2008hs} or radial flow~\cite{PhysRevLett.109.102301}. The latter is characteristic to hydrodynamical expansion of the system and its presence might require a fireball in local thermodynamical (kinetic) equilibrium. Similar dynamics observed in smaller systems such as pp or p-Pb, where hydrodynamics was assumed to be not applicable due to the absence of a QGP phase, can be explained by certain QCD effects like color reconnection~\cite{Sjostrand:2007gs, DerradideSouza:2016kfn}.

On the other hand, increased abundances of strange hadrons in heavy-ion collisions relative to that in pp was originally proposed in 1982 as a signature of QGP~\cite{Rafelski:1982pu} and was first observed in Pb-Pb collisions at SPS~\cite{Andersen:1999ym}.
Alternatively, in statistical hadronization models~\cite{BraunMunzinger:2001ip} the observed strange particle abundances across collision systems can be explained as a canonical suppression of strange quark production in pp collisions, which then gradually subsides for larger system sizes~\cite{Vislavicius:2016rwi}.

To understand how important the initial system configuration are for the final state observables, one would study pp, p-Pb and Pb-Pb collisions. So far, changing the colliding system does not seem to modify relative particle abundances provided that event activities are similar. Now, by comparing the most recent data from pp collisions at \s{13} to that at lower energies, we can isolate the center-of-mass energy dependence of hadrochemistry and kinematics.


\section{Analysis and results}

The analysis of a 50M minimum bias (MB) triggered event sample of pp collisions at \s{13}, recorded by ALICE~\cite{Aamodt:2008zz} in 2015, has lead to the measurements of the production of $\pi$, $K$, $p$, strange and multi-strange particles. A hit in either V0 scintillators or in the SPD in coincidence with signals from beam pick-up counters was used for MB triggering and events containing more than one primary vertex within $|z|<10~\textrm{cm}$ were discarded as pileup. Acceptance and efficiency corrections were calculated from simulations, using PYTHIA8 (Monash-2013 tune)~\cite{Sjostrand:2007gs} as particle generator and GEANT3 for describing particle transport in the ALICE detector. In addition, the production of strange hadrons has been studied as a function of the event activity, characterized by the average charged particle multiplicity \dndeta\ measured at mid-rapidity ($|\eta|<0.5$). To avoid auto-correlations, event activity classes were selected using signals in the V0 detector -- two scintillator arrays covering $-3.7 < \eta < -1.7$ and $2.8 < \eta < 5.1$~\cite{Aamodt:2008zz}.

Charged pions, kaons and protons were identified in the ALICE central barrel following the approach used in pp collisions at \s{7}~\cite{Adam:2015qaa}. The (multi-)strange baryons and $K^{0}_{S}$ were reconstructed using daughter tracks from the weak decays in the rapidity window $|y|<0.5$.

\begin{figure}[!h]
  \begin{center}
    \includegraphics[width=.9\textwidth]{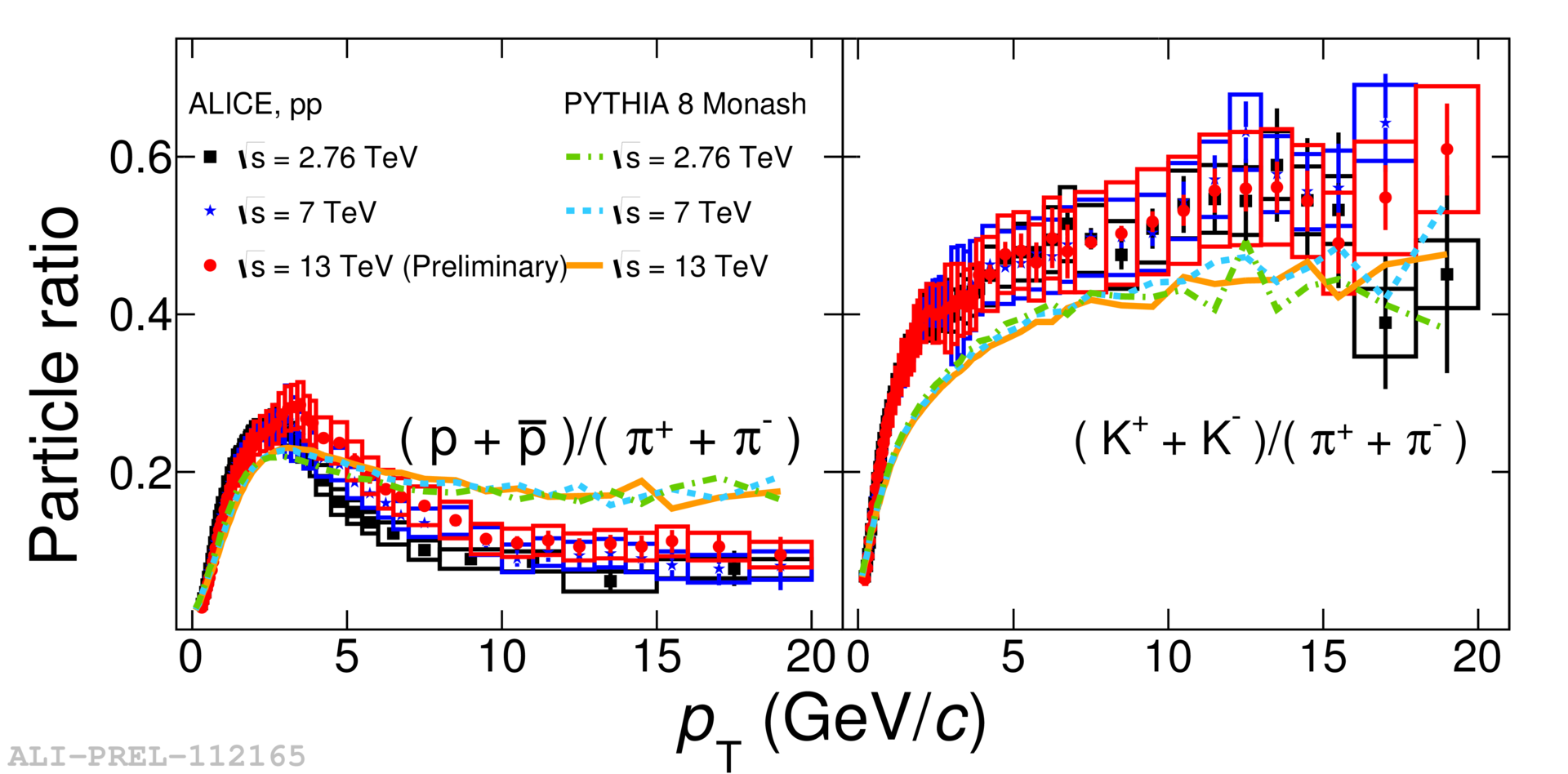}
    \caption{\pT-differential $p/\pi$ (left) and $K/\pi$ (right) ratios measured at different \so\ with comparison to PYTHIA8 predictions.}
    \label{fig:pandkoverpi}
  \end{center}
\end{figure}

The \pT-differential $p/\pi$ and $K/\pi$ ratios measured in a rapidity window $|y|<0.5$ in MB pp collisions at \s{13} are shown in \fig{pandkoverpi}, along with similar measurements at \so\ ~2.76 and 7~TeV. While there is no significant evolution of $K/\pi$ ratios with \so, the peak of $p/\pi$ ratio shifts to slightly higher values of \pT\ with the increase of \so. Note that a minor modification of baryon-to-meson ratio is expected considering a small increase in \dndeta\ with \so~\cite{Adam:2015pza}. A comparison to PYTHIA8 predictions reveals that not only $K/\pi$ and $p/\pi$ ratios are not described, but also the evolution of $p/\pi$ ratio with \so\ is not captured within the generator framework.

\begin{figure}[!h]
  \begin{center}
    \includegraphics[width=.49\textwidth]{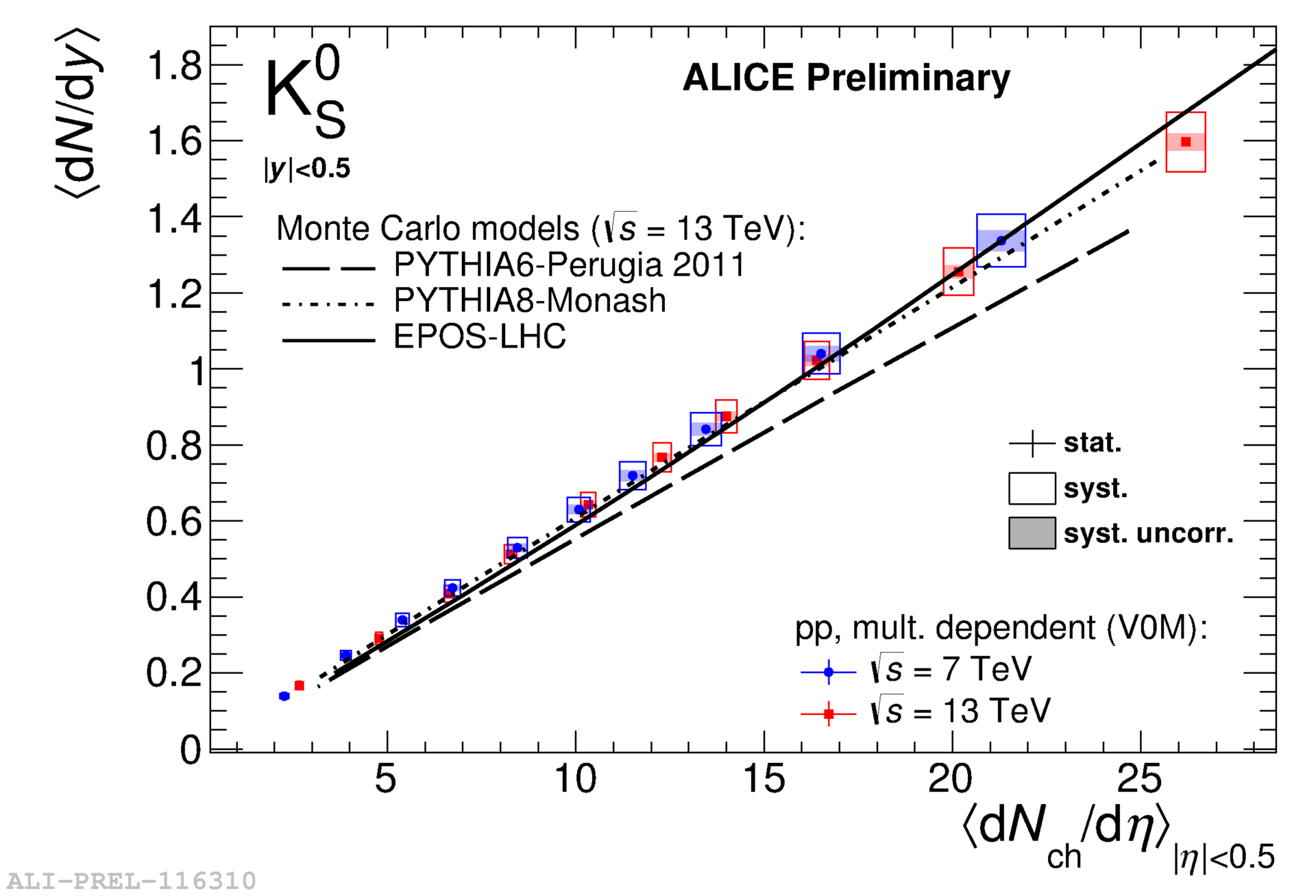}
    \includegraphics[width=.49\textwidth]{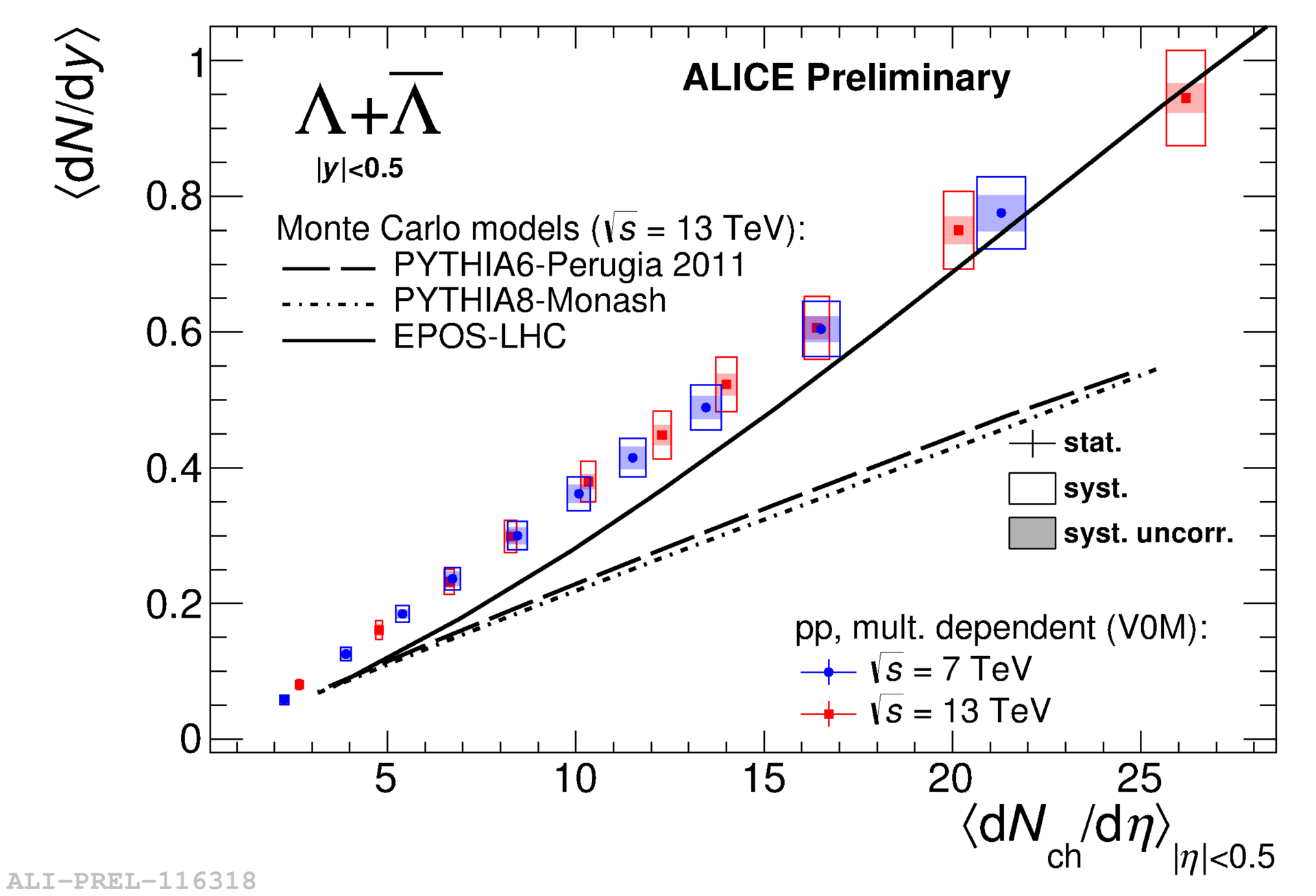}
    \includegraphics[width=.49\textwidth]{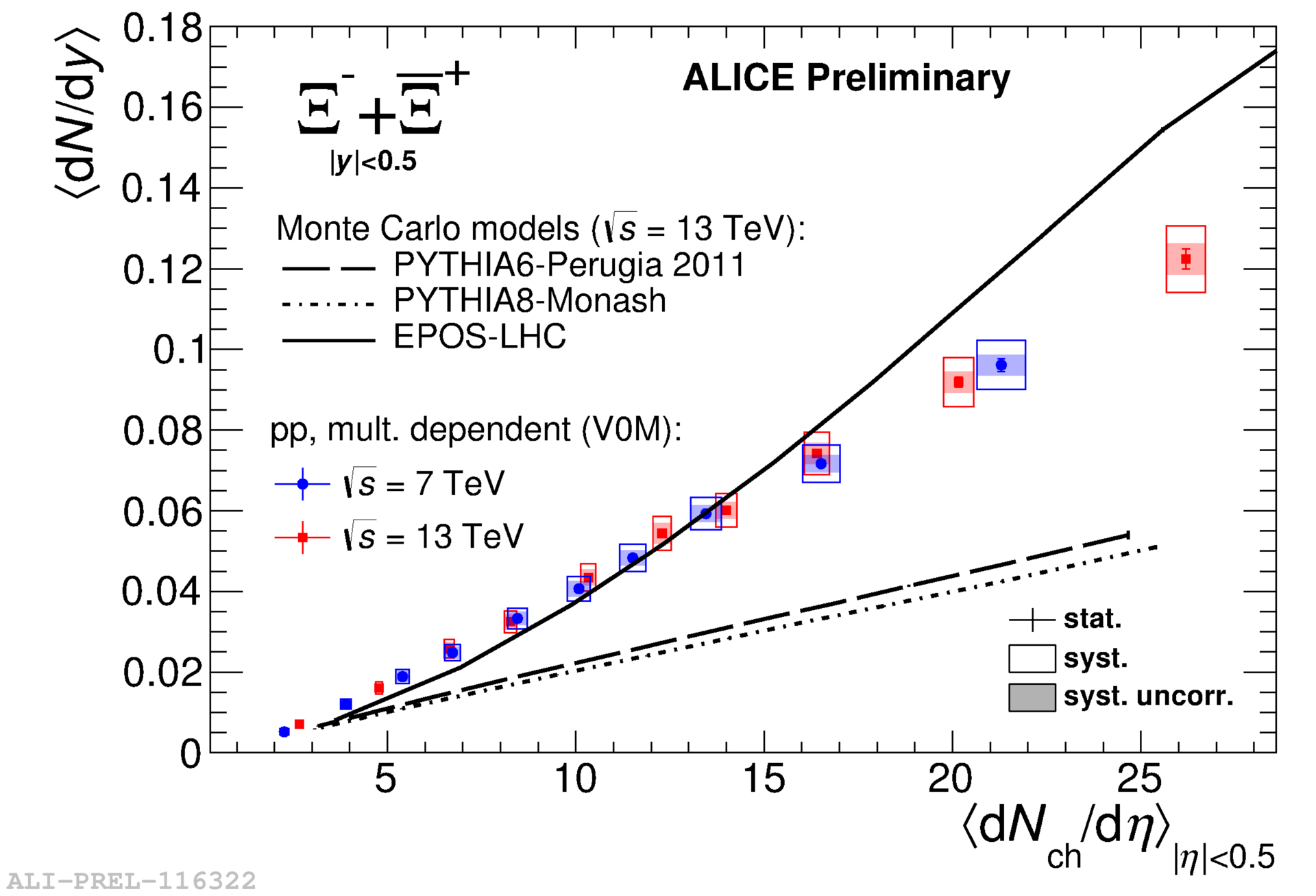}
    \includegraphics[width=.49\textwidth]{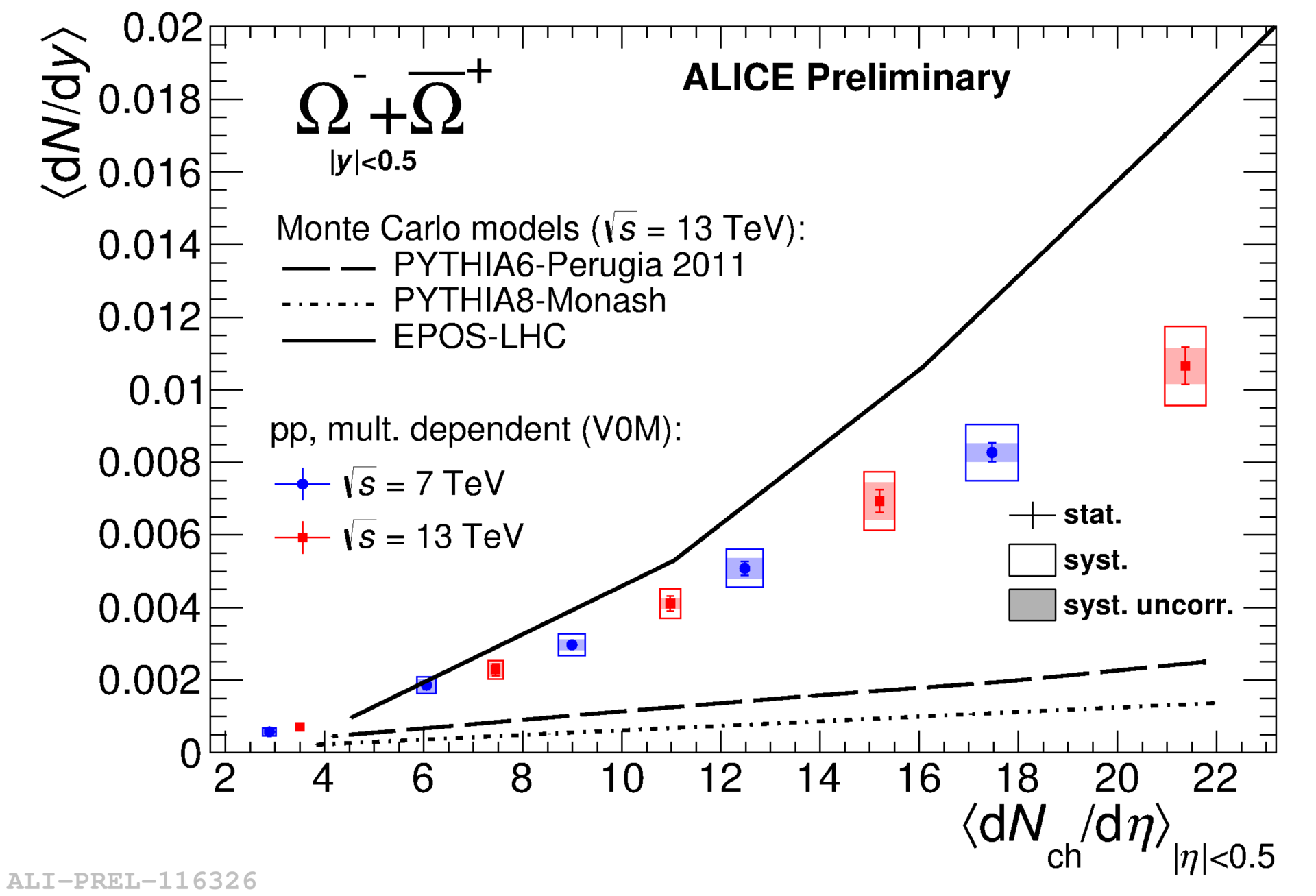}
    \caption{\pT-integrated $K^{0}_{S}$, $\Lambda+\overbar{\Lambda}$, $\Xi^{-}+\overbar{\Xi}^{+}$ and $\Omega^{-}+\overbar{\Omega}^{+}$ yields as a function of charged particle multiplicity at $|\eta|<0.5$ measured in pp collisions at \so~=~7 (red) and 13 (blue) TeV with comparison to EPOS-LHC~\cite{Pierog:2013ria} and PYTHIA6/PYTHIA8~\cite{Sjostrand:2007gs} predictions.}
    \label{fig:13TeVYields}
  \end{center}
\end{figure}

The \pT-integrated proton- and hyperon-to-pion ratios as a function of center-of-mass energy have previously been shown in~\cite{DerradideSouza:2016kfn}. While $p/\pi$ ratios saturate at LHC energies, $\Xi/\pi$ and $\Omega/\pi$ ratios exhibit hints of an increase between MB pp collisions at \so~=~7 and 13 TeV. To further investigate this enhancement, a comparison of \pT-integrated $K^{0}_{s}$, $\Lambda + \overbar{\Lambda}$, $\Xi^{-}+\overbar{\Xi}^{+}$ and $\Omega^{-}+\overbar{\Omega}^{+}$ yields in pp collisions at \so~=~7 and 13 TeV as a function of \dndeta\ is shown in~\fig{13TeVYields}. We observe similar particle abundances at similar final state multiplicities for the two different center-of-mass energies, indicating that particle production is dominantly driven by the event activity and not by \so. The increase of yields with \dndeta\ is stronger for hadrons with larger strangeness content, and given the saturation of $p/\pi$~\cite{Adam:2016emw, DerradideSouza:2016kfn}, it indicates that this effect is related to strangeness enhancement (suppression) in large (small) systems and not to the baryonic number. A comparison to Monte Carlo predictions shows that the existing generators do not capture the evolution of (multi-) strange hadron yields with \dndeta: while both PYTHIA6/PYTHIA8~\cite{Sjostrand:2007gs} and EPOS-LHC~\cite{Pierog:2013ria} describe $K^{0}_{s}$ yields well, discrepancies between model predictions and data grow for baryons with larger strangeness content.

\begin{figure}[!h]
  \begin{center}
    \includegraphics[width=.49\textwidth]{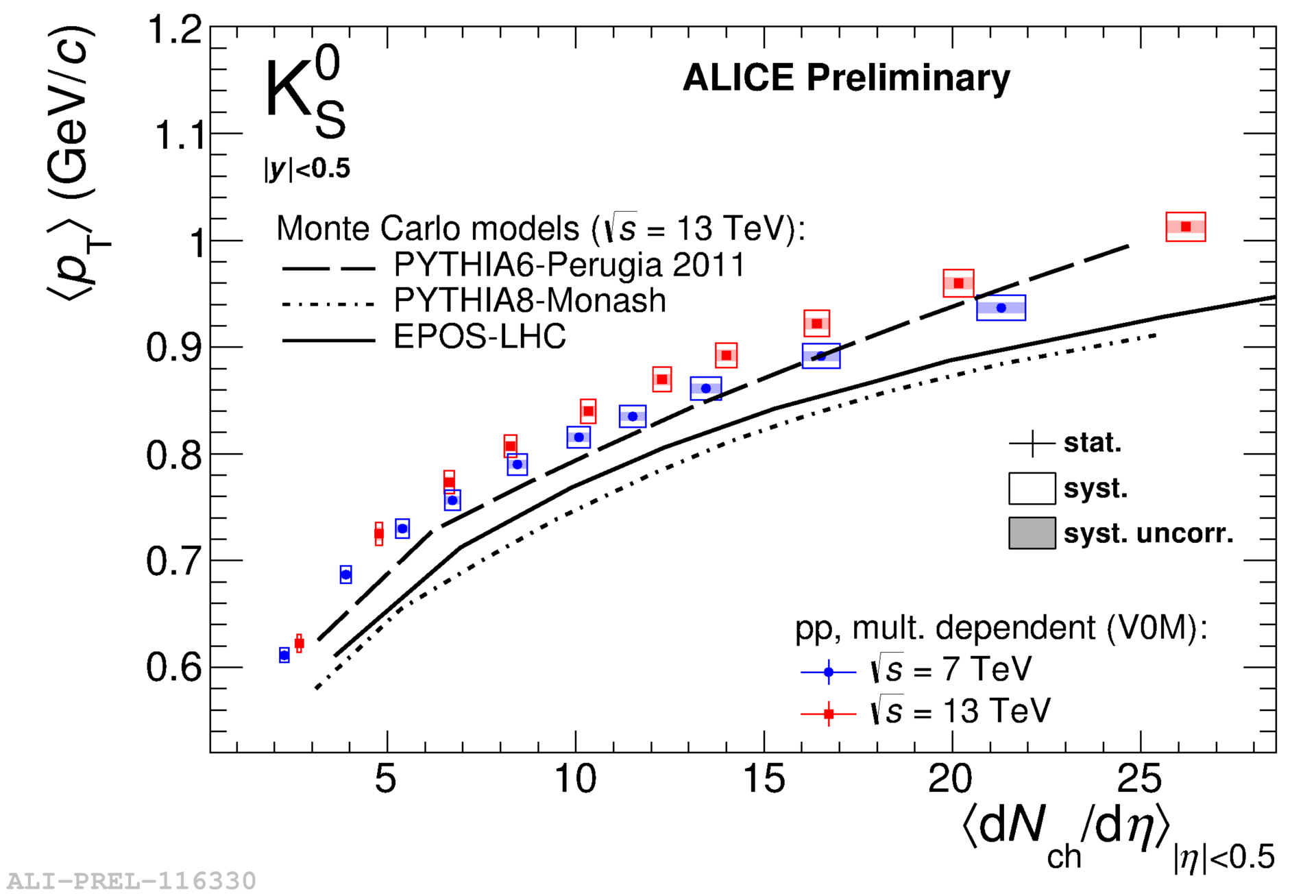}
    \includegraphics[width=.49\textwidth]{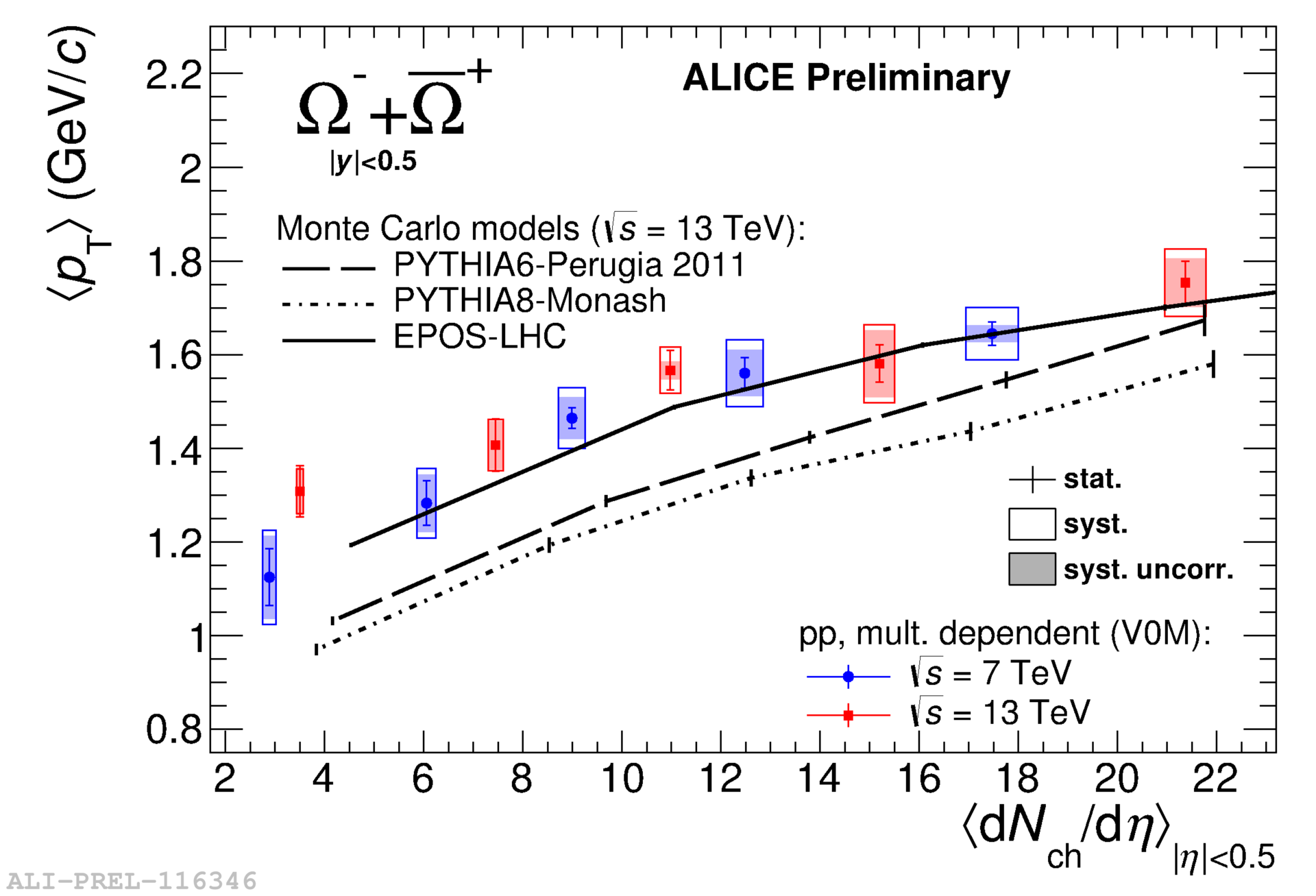}
    \caption{$\langle\pT\rangle$ as a function of multiplicity for $K^{0}_{s}$ (left) and $\Omega^{-} + \overbar{\Omega}^{+}$ (right) measured in pp collisions at \so~=~7 (blue) and 13~TeV (red) with comparison to MC predictions.}
    \label{fig:MeanPt}
  \end{center}
\end{figure}

The mean transverse momentum \mpt\ of $K^{0}_{s}$ and $\Omega^{-}+\overbar{\Omega}^{+}$ as a function of multiplicity measured in pp collisions at \so~=~7 and 13 TeV is shown in \fig{MeanPt}. The MC models predict a hardening of the spectra with multiplicity, which is observed in data. However, the rate of hardening is not predicted correctly. We also observe a small increase of $K^{0}_{s}$ \mpt\ at higher \so\ for similar final state multiplicities. Whether the same trend is observed in case of (multi-) strange baryons is not clear due to the present systematic uncertainties, but similar behavior has previously been reported at lower energies for charged particles~\cite{Abelev:2013bla}.
\clearpage

\section{Summary}

The ALICE collaboration has measured and reported results on light flavor particle production as a function of multiplicity in pp collisions. To isolate the impact of \so\ on final state observables, measurements in pp were performed for two different center-of-mass energies, 7 and 13 TeV. We observe a small blueshift of the maximum in \pT-differential $p/\pi$ spectra ratio at \s{13} as compared to lower energies, while no evolution is seen in $K/\pi$. The \pT-integrated $p/\pi$ ratios saturate at LHC energies, while hyperon-over-pion ratios hint towards a small increase between \so~=7 and 13~TeV. The integrated hadron yields show a very good scaling behavior with event activity and are very similar at comparable \dndeta\ for different collision energies. On the other hand, \mpt\ of $K^{0}_{s}$ exhibits an increase in \s{13} pp collisions as compared to 7~TeV. This indicates that the hadrochemistry is dominantly driven by \dndeta, even though the dynamics of particle production might be different at different energies. Finally, the most common tunes of MC generators do not provide a satisfactory description of the evolution of these observables with multiplicity.






\bibliographystyle{elsarticle-num}
\bibliography{MyBibliography}

\begin{thebibliography}{10}
\expandafter\ifx\csname url\endcsname\relax
  \def\url#1{\texttt{#1}}\fi
\expandafter\ifx\csname urlprefix\endcsname\relax\def\urlprefix{URL }\fi
\expandafter\ifx\csname href\endcsname\relax
  \def\href#1#2{#2} \def\path#1{#1}\fi

\bibitem{Aamodt:2008zz}
K.~Aamodt, et~al. (ALICE Collaboration), JINST 3 (2008)
  S08002.
\bibitem{Abelev:2013haa}
B.~B. Abelev, et~al. (ALICE Collaboration), Phys. Lett.
  B728 (2014) 25--38.
\bibitem{Adam:2016emw}
J.~Adam, et~al.(ALICE Collaboration), arXiv:1606.07424.

\bibitem{Fries:2008hs}
R.~J. Fries, V.~Greco, P.~Sorensen, Ann. Rev. Nucl. Part. Sci. 58 (2008) 177--205.

\bibitem{PhysRevLett.109.102301}
K.~Werner, Phys. Rev. Lett.
  109 (2012) 102301.

\bibitem{Sjostrand:2007gs}
T.~Sjostrand, S.~Mrenna, P.~Z. Skands,
  Comput. Phys. Commun. 178 (2008) 852--867.

\bibitem{DerradideSouza:2016kfn}
R.~Derradi~de Souza (ALICE Collaboration), J. Phys. Conf. Ser. 779~(1)
  (2017) 012071.

\bibitem{Rafelski:1982pu}
J.~Rafelski, B.~Muller,
  Phys. Rev. Lett. 48 (1982) 1066.

\bibitem{Andersen:1999ym}
E.~Andersen, et~al., Phys. Lett. B449 (1999) 401--406.

\bibitem{BraunMunzinger:2001ip}
P.~Braun-Munzinger, D.~Magestro, K.~Redlich, J.~Stachel, Phys. Lett. B518 (2001) 41--46.

\bibitem{Vislavicius:2016rwi}
V.~Vislavicius, A.~Kalweit, arXiv:1610.03001

\bibitem{Adam:2015qaa}
J.~Adam, et~al. (ALICE Collaboration), Eur. Phys. J. C75~(5)
  (2015) 226.

\bibitem{Adam:2015pza}
J.~Adam, et~al. (ALICE Collaboration), Phys.
  Lett. B753 (2016) 319--329.

\bibitem{Pierog:2013ria}
T.~Pierog, I.~Karpenko, J.~M. Katzy, E.~Yatsenko, K.~Werner, Phys. Rev. C92~(3) (2015) 034906.

\bibitem{Abelev:2013bla}
B.~B. Abelev, et~al. (ALICE Collaboration), Phys. Lett. B727
  (2013) 371--380.

\end{thebibliography}





\end{document}